\documentclass[aps,pra, showpacs,onecolumn,superscriptaddress,twoside,floatfix,a4]{revtex4}
\usepackage{bbm,bm}
\usepackage{amsmath, amssymb}
\usepackage{color}
\usepackage{graphicx,epsfig}
\usepackage{pifont}

\input epsf.sty

\def\ket#1{\mathinner{|{#1}\rangle}}
\def\bra#1{\mathinner{\langle{#1}|}}

\newcommand{\tick}{\ding{51}}
\newcommand{\notick}{\ding{53}}

\usepackage{graphicx,indentfirst,amsmath,subfigure,psfrag}

\begin{document}

\title{Secure Multi-Party Computation with a Dishonest Majority via Quantum Means}

\author{Klearchos Loukopoulos}
\affiliation{Department of Materials, University of Oxford, Parks Road, Oxford OX1 4PH}
\email{klearchos.loukopoulos@seh.ox.ac.uk}
\author{Daniel E. Browne}
\affiliation{Department of Physics and Astronomy, University College London, Gower Street, London WC1E 6BT, United Kingdom.}

\pacs{03.67.Dd,03.65.Ud}

\begin{abstract}

 We introduce a scheme for secure multi-party computation utilising the quantum correlations of entangled states.  First we present a scheme for two-party computation, exploiting  the correlations of a Greenberger-Horne-Zeilinger state to provide, with the help of a third party, a near-private computation scheme. We then present a variation of this scheme which is passively secure with threshold $t=2$, in other words, remaining secure when pairs of players conspire together provided they faithfully follow the protocol. Furthermore we show that the passively secure variant can be modified to be secure when cheating parties are allowed to deviate from the protocol. We show that this can be generalised to computations of $n$-party polynomials of degree 2 with a threshold of $n-1$. The threshold achieved is significantly higher than the best known classical threshold, which satisfies the bound  $t<n/2$. Our schemes, each complying with a different definition of security, shed light on how Lo's seminal theorem translates into a measurement based scheme and highlight which physical assumptions are necessary in order to achieve quantum secure multi-party computation.
\end{abstract}

\maketitle

\section{Introduction}

Secure multi-party computation (SMPC) is an important and well-studied cryptographic protocol. It was originally introduced by Yao \cite{yaomil} in the form of the ``millionaire problem'', in which two millionaires wish to discover which of them is the richest without revealing the size of their personal fortunes. In its general form, SMPC refers to the case where $n$ parties, each holding a set of private variables, want to compute a publicly available function $f$ without revealing any information about their variables to other parties, beyond that revealed by the output of function itself.  Real world applications of secure multi-party computation include market clearing price scenarios, secure voting and on-line bidding \cite{smpcapps}.

It would be simple to achieve SMPC if a trusted third party were available. In this ``ideal scenario'', each party would securely send their private data to this trusted third party, who would perform the computation privately and then announce the result.  Studies of SMPC are therefore concerned with scenarios where no party can be trusted. Such schemes are deemed to be secure (under specified constraints on the computational power and activities of the parties) when the information learned by each party about their neighbours' inputs matches the ideal case.


 Secure multi-party computation has been studied under a variety of assumptions. These may be limitations on the computational power available to adversaries or restrictions to the level to which parties are allowed to deviate from an agreed protocol. A protocol is called ``computationally secure'' when its  security assumes that an adversary lacks the computational power needed to perform a certain computation, which is believed to be untractable. The ability to compute this function would allow him to break the security of the protocol. An example of this kind of security is the well-known RSA public key system, which bases its security on the hardness of factoring large numbers.  The downside to computational security is that the schemes may be vulnerable to new algorithms and new technologies, such as quantum computation, where an efficient factoring algorithm \cite{factorprimes} is known.
  A protocol is  called  ``information theoretically'' secure when its security properties hold independent of the computational power of the adversary. An example of an information-theoretically secure cryptographic system is the well-known Vernam cipher \cite{vernam}.

A further ingredient in security assumptions of an SMPC scheme is the allowed behaviour of the participants in the protocol.  This has been an important component in classical studies of the SMPC computation, but has so far, not been focussed upon in quantum approaches. The most important behaviour models occuring in the classical literature are the ``passive'' security model, and the ``active'' model.

A protocol has ``passive security'' with ``threshold'' $t$, if it remains secure provided all parties follow the protocol exactly, but $t$ or fewer parties are ``corrupted''. The corrupted parties may form a coalition,  sharing data during the execution of the protocol. A protocol is ``actively secure'' with threshold $t$ if security is retained when parties are allowed to deviate arbitrarily from the protocol and $t$ or fewer parties are corrupted. In particular, parties in an actively secure model may send incorrect data during the protocol in an attempt to trick other parties into revealing extra information about their inputs. In addition to these standard definitions, we shall call a protocol ``private'' if security is achieved when all parties follow the protocol and do not share data (equivalent to the $t=1$ passive case). We call a protocol ``nearly private'' when less information is revealed about the parties inputs than public computation but the ideal scenario is not quite attained.
 
The first solutions to the SMPC problem were based on computational security assumptions. These include Yao's solution to the millionaire's problem \cite{yaomil} and more general treatments \cite{GMW87,CDG87}. Later, information theoretic solutions were shown in ~\cite{BGW88,CCD88,RB89,Bea89,DamsNiel}. A summary of the assumptions, thresholds and efficiency of these protocols can be found in \cite{DamsNiel}. The best thresholds for these schemes are upper bounded by $n/2$ - in other words, an honest majority is required.
 
 
After the success of quantum key distribution \cite{qkd1,qkd2,qkd3}, there was a concerted attempt to construct protocols with quantum enhancement for a number of key cryptographic primitives, such as bit commitment \cite{qbitcommit1,qbitcommit2} and secure multi-party computation. Bit commitment was shown to be impossible \cite{qbitcommit1, qbitcommit2}, however, quantum protocols were successfully found for quantum secret sharing of classical information \cite{qsecshare1, qsecshare2, qsecshare3, qsecshare4} and quantum information \cite{qsecshare5,qsecshare6}. 

It was thus natural to consider whether quantum advantages may assist in the problem of secure multi-party computation. While computational security is possible in the quantum case, e.g. by combining trapdoor functions together with the detectable byzantine agreement protocol proposed by Fitzi, Gottesman et. al. in \cite{fitzigott}, surprisingly, this advantage was found to be limited for the case of unconditional security. In fact, it was  shown by Lo \cite{Lo} (see \cite{colbeck,salvail} for recent generalisations) that  deterministic  two-party setting computation was impossible, even with quantum means (see \cite{colbeck,salvail} for recent generalisations of this result).  \cite{footnote1}

 Here, we uncover key assumptions on Lo's seminal theorem through the use of several security models and show which parts of the theorem correspond to different security assumptions. We identify the most general security model which provides unconditional security while remaining compliant with Lo's theorem and construct a protocol which satisfies this model.


 We introduce a ``quantum passive'' security model, a variant of the passive security model well-studied in the classical case. A key assumption in Lo's and Colbeck's no-go theorems is that the entire protocol may be modeled by a ``unitary black box". In the quantum passive model,  this assumption doesn't hold. Under this model, we offer a quantum solution to the SMPC problem, by presenting a protocol which is secure against external eavesdroppers and coalitions between party members within the quantum passive secure model which we introduce.  Furthermore, we introduce a No Quantum Cheating Channel (NQCC) model, which allows corrupted parties to lie or deviate from the protocol and prove that our protocol offers a solution to the SMPC problem compliant with NQCC security.

 Our schemes exploit the non-classical correlations of Greenberger-Horne-Zeilinger (GHZ) states \cite{ghz1, ghz2, ghz3, ghz4, ghz5}, recently shown \cite{danclascom} to be resources for classical computation similar to the way cluster states are a resource for universal quantum computation \cite{onewayqc1,onewayqc2}. The advantage provided by our quantum schemes is that, for certain functions, it provides a higher security threshold than all current classical schemes, remaining secure even when  all but one parties are dishonest.

  The structure of this paper is as follows, in section ~\ref{quantumnand} we briefly review Anders and Browne's reinterpretation of the well-known GHZ quantum correlation in terms of the computation of the Boolean AND-function.
   In section~\ref{secprot} we show how this may be developed into a nearly private multi-party computation scheme. In section~\ref{passecan}, we perform a privacy analysis on the protocol of section ~\ref{secprot} and discuss its weaknesses.
 Section ~\ref{modelpassec} contains our definition for our quantum security model and in section~\ref{passsecprot}, we then present a variation of the same protocol which is passively secure with threshold $t=2$. Section~\ref{actsecan} contains a passive security analysis of our protocol and a proof that it is passively secure, which includes a discussion on how attacks similar to those in \cite{colbeck,Lo} relate to our scheme and the quantum security model which we introduce. Then, in section~\ref{nqccmodel} we define the No Quantum Cheating Channel Model (NQCC), which is a variation of the passive model where corrupted parties can lie. Section ~\ref{nqccprot} offers an upgrade to the passively secure protocol, so that it is NQCC-secure and after this, in section~\ref{activesecuran} we present its security analysis.
   %
   %
    In section~\ref{betterthresh} we demonstrate that quantum mechanics has the potential to offer a better corruption threshold than classical protocols, indeed for certain function classes, it can become maximal, that is $t=n-1$ for an $n$-member party.

\section{Greenberger-Horne-Zeilinger correlations for distributed computation and secret sharing}
\label{quantumnand}

 In this section we briefly review the recent work \cite{danclascom}, in which the correlations present in measurements upon the Greenberger-Horne-Zeilinger (GHZ) state are interpreted as a distributed computation of the Boolean AND-function. This can be seen most clearly by considering the stabilizer equations for the GHZ state, first presented by Mermin \cite{ghz4},
%
%
\begin{equation}
\begin{split}
& \sigma_z \otimes \sigma_z \otimes \sigma_z \ket{\psi} = \ket{\psi}, \\
& \sigma_z \otimes \sigma_x \otimes \sigma_x \ket{\psi} = \ket{\psi}, \\
& \sigma_x \otimes \sigma_z \otimes \sigma_x \ket{\psi} = \ket{\psi}, \\
& \sigma_x \otimes \sigma_x \otimes \sigma_z \ket{\psi} = -\ket{\psi}, \\
\end{split}
\label{eigeneq}
\end{equation}
where for notational convenience later in this paper, we have chosen the locally equivalent GHZ state $\ket{\psi} =(1/\sqrt{2})  (\ket{y_-y_-y_+}+\ket{y_+y_+y_-})$, with $\ket{y_+} =(1/\sqrt{2}) (\ket{0}+i\ket{1})$ and $ \ket{y_-} = (1/\sqrt{2}) (\ket{0}-i\ket{1})$. This is locally equivalent to the more well-known GHZ state $(1/\sqrt{2})  (\ket{000}+\ket{111})$ and inability for these equations to be simultaneously satisfied by c-number scalar values, representing the measured value in a hidden variable theory, is sometimes known as the GHZ paradox.

We imagine that the three qubits are divided among three parties, each of which will measure in either the $\sigma_{z}$ or $\sigma_{x}$ basis and label these measurement operators $O_{0}=\sigma_{z}$ and $O_{1}=\sigma_{x}$. We can then rewrite the four equations above in compact form.
\begin{equation}
O_{a}\otimes O_{b}\otimes O_{a\oplus b}\ket{\psi}=(-1)^{\textrm{AND}(a,b)}  \ket{\psi}
\end{equation}
where $a$ and $b$ are bit-values and $\oplus$ denotes addition modulo 2.

Note that the value of the Boolean AND of bits $a$ and $b$ is encoded in the eigenvalues of these equations. Representing the measured eigenvalues $+$ and $-$ with the bit values $M_{i} \in\{0,1\}$ we see  
\begin{equation}
M_1\oplus M_2 \oplus M_3 = \textrm{\textrm{AND}(a,b)}
\label{calcnand}
\end{equation}

We see therefore see that if three parties sharing $\ket{\psi}$ make measurements determined by bit-values $a$, $b$ and $a\oplus b$, the parity of their output bits is equal to $\textrm{AND}(a,b)$.
%
%
%
%
%
 An interesting aspect of this is that the computation can be done in a distributed manner. The qubits which form the GHZ state do not have to be in the same spacial point, they can be distributed between spatially separated parties. The outcome of the computation is naturally encoded in the parity of bits held by the three parties.  This is a simple form of secret sharing \cite{secretshare} because the value is only revealed if the three parties share their data. 
  In the next sections, we use this property as the basis of protocol for secure multi-party computation.


\section{Scheme A: A nearly private multi-party computation protocol}
\label{secprot}

\begin{figure}[htb!]
\includegraphics[width=3in]{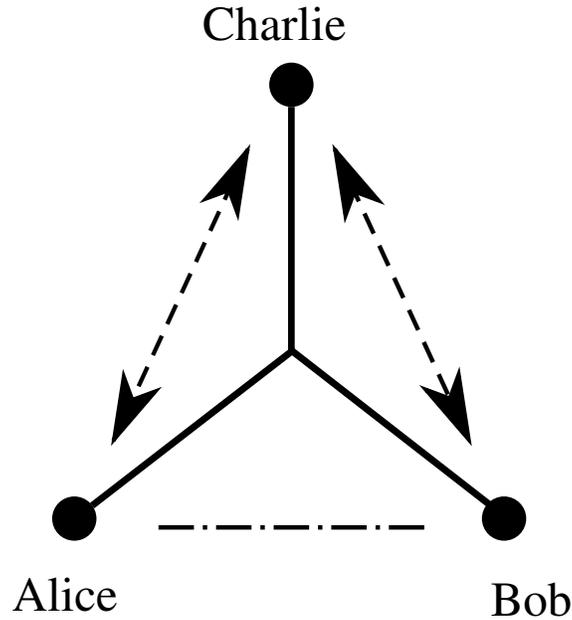}

\caption{ In this figure we present graphically our scheme. The three party members, Alice, Bob and Charlie are connected with straight lines which represent the GHZ state. The dashed line corresponds to the shared randomness resource (e.g. a Bell state) which is shared between Alice and Bob and the dashed lines, whose ends are arrows, that exist between Charlie and Alice, Bob represent a secure classical channel.}
\label{3partyset}
\end{figure}

 In this section,  we introduce a scheme for  multi-party computation between two parties, Alice and Bob. To circumvent Lo's \cite{Lo} no-go theorem a third party, Charlie, is required. Adding a third party allows a measurement-based scheme to be employed and this introduces an irreducibly classical part, the measurement outcomes into the protocol. It is this classical part which makes the computation model differ to the ``unitary black-box" model used in no-go theorems \cite{colbeck,Lo}.The scheme has enhanced privacy compared to public computation, but Charlie learns more information about Alice and Bob's input than in the ideal scenario. We therefore call this scheme ``nearly private''. To implement the scheme Alice, Bob and Charlie must share a GHZ state $\ket{\psi}$ and additionally, each pair must share secret correlated random bits. In addition, Charlie is able to send data on a secure classical channel to Alice and also to Bob, as shown in Fig.~\ref{3partyset}. The correlated private bits and the secure channel can both be achieved in information theoretically secure manner by standard quantum key distribution protocols \cite{qkd1,qkd2,qkd3}.

%


Let $f\left(x_1,\ldots, x_n, y_1,\ldots, y_n\right)$  be the function to be calculated and let $\vec{x}=(x_1,\ldots, x_n)$ and $\vec{y}=( y_1,\ldots y_n)$ represent Alice and Bob's input data. 
 In order to simplify our protocol we will make use of the fact that any Boolean function $f\left( \vec{x},\vec{y}\right): {\{0,1\}}^n\times {\{0,1\}}^n \rightarrow {\{0,1\}}, \vec{x},\vec{y} \in {\{0,1\}}^n$ can be calculated as the inner product of two vectors of polynomials,  $P_i\left(\vec{x}\right)$ and $Q_i\left(\vec{y}\right)$ \cite{vandamssnloc}
%
%
\begin{equation}
f\left(x_1,..,x_n, y_1,..,y_n\right) = \bigoplus_{i=1}^m P_i\left(\vec{x}\right)Q_i\left(\vec{y}\right),
\label{vandam}
\end{equation}
\noindent where the sum operator corresponds to addition modulo $2$. The number of terms that will be needed, $m$, is at worst bounded by $2^n$, where $n$ is the length of the input vectors, therefore this decomposition can only be practically employed when $m$ scales polynomially with $n$, $m \backsim \text{poly}\left(n \right)$.  
 The function can be evaluated by the calculation of each product $P_i Q_i$ in turn. The polynomials $P_{i}$ and $Q_{i}$ can be calculated locally, and hence privately by Alice and Bob respectively. All that is required in addition is the ability to compute $P_{i}Q_{i}=\textrm{AND}(P_i,Q_i)$ for each value of $i$. Our first protocol exploits the correlations of a GHZ state to achieve this.

The protocol proceeds in the following steps, all addition is performed modulo 2:

\begin{enumerate}
\item Repeat steps $2$-$7$ for all the terms present in equation  \eqref{vandam}, starting with $i=1$.
\item Alice and Bob calculate $P_{i}$ and $Q_{i}$ locally.
\item Alice and Bob generate a private shared random bit $r_{i}$, e.g. by suitable measurements on a Bell state, or via a secure communications channel.
\item Alice transmits to Charlie the bit-value $P_{i}\oplus r_{i}$.
\item Bob transmits to Charlie the bit-value $Q_{i}\oplus r_{i}$.
\item Charlie adds together these bits to reconstruct $P_{i}\oplus Q_{i}$ = $\left(P_{i} \oplus r_i \right) \oplus \left(Q_{i} \oplus r_i \right) $. 
\item Alice, Bob and Charlie measure their qubits of the GHZ state $\ket{\psi}$ as determined by bit values $P_{i}$, $Q_{i}$ and $P_{i}\oplus Q_{i}$, for bit-value 0, they measure $\sigma_{z}$ and for bit-value 1, they measure $\sigma_{x}$.
\item Once this has been completed for all $i$ terms, Alice and Bob sum their local measurement outcomes. They send their summation bits to Charlie, who sums them with his own measured outcomes.
\item Charlie reveals the value of $f(\vec{x},\vec{y})$.
\end{enumerate}




The correctness of the protocol follows simply from the analysis in the previous section. Due to the correlations of the GHZ state, the value of $P_{i}Q_{i}$ is encoded in shared secret form across the three parties. We analyse the privacy in the following section.

\section{Privacy Analysis of Scheme A} \label{passecan}

In this section we shall examine Scheme A, step by step, and identify how much each party learns about the inputs of Alice and Bob at each stage. 

In \textit{steps 1 to 3} all operations are local, therefore no information about the private bits can be obtained. After \textit{steps 4 to 6}, Charlie receives the parity of Alice and Bob's private bits $P_{i}$ and $Q_{i}$. By use of private channels, no third party could learn these bit values. By use of random bit $r$, Charlie does not learn anything about the individual values $P_{i}$ and $Q_{i}$ other than their parity. 
In \textit{steps 7 to 9} the three parties exchange bits which, individually, carry no information about either the input or outputs of the function. After the value of the function is announced, Alice has learnt nothing about Bob's inputs more than she would in  the ideal scenario. Bob has learnt a similar amount about Alice's inputs. 

The only deviation from the ideal scenario is the parity information for each term learnt by Charlie. This information could, under certain circumstances, be used by Charlie to reconstruct some of Alice and Bob's input data. As an extreme example of this, consider the two-input function $f(x_{1},y_{1})=x_{1}y_{1}$. If the function outcome is $0$ and the parity of the bits is even, Charlie knows with certainty  that both Alice and Bob's inputs were $0$.
We therefore say this protocol is nearly but not completely private.

Since Charlie knows the parity of Alice and Bob's $P_{i}$ and $Q_{i}$ bits at every stage the protocol is intrinsically insecure against any coalition. If Charlie and Bob collaborate, for example, they can learn all of Alice's $P_{i}$ bits. The protocol, and indeed any protocol where Charlie learns similar parity information,  is therefore not passively secure above a trivial threshold $t=1$. 
In the section ~\ref{passsecprot}, we modify the above scheme such that Charlie never learns such parity information, and in doing so, introduce a scheme which is passively secure.

\section{Model for passive security}
\label{modelpassec}

Passive security is an important security model in classical SMPC. In this section we introduce a variant ``quantum passive security'', a generalisation of the classical model, where the participants behaviour with respect to quantum resources is specified.

 In classical passive security, which we again summarise below, corrupted parties can exploit the information they gain throughout the execution of an SMPC protocol, even collaboratively by forming coalitions but do not deviate from the protocol. In classical SMPC this means that corrupted parties can exchange classical information and use the total information they infer to learn the private data of honest parties, this approach is also known as ``honest but curious".
 
 In our quantum passive security model, similarly to the classical models, corrupted adversaries are allowed to exchange classical information however,  they are not permitted to exchange quantum data, and do not posses any shared quantum resources additional to the GHZ states provided through the protocol. In general we shall consider,  SMPC protocols that are $n$-sided, that is, all parties learn the computation outcome, however,  we discuss a $1$-sided variation of our protocol in section ~\ref{actsecan} to show that it parries so-called EPR attacks.

The above are summarized in the following table,

\begin{center}
  \begin{tabular}{ | c || c | c |}
    \hline
    Property & Classical passive security & Quantum passive security \\ \hline
    Private data are not inferred by corrupted parties & \tick & \tick \\ \hline
    Corrupted parties do not lie - their outputs are true to the protocol & \tick & \tick \\ \hline
    Corrupted parties exchange classical information & \tick & \tick \\ \hline
    Corrupted parties exchange quantum information & N/A & \notick \\ \hline
 
  \end{tabular}
\end{center} 
 
 The restriction on the exchange of quantum information is important. It is this assumption which means that Lo and Colbeck's unitary black box model (ad hence their no-go theorems) does not apply. Below, we shall explain how allowing such communication does indeed break the presented protocol.
 


\section{Scheme B: Passively secure multi-party computation protocol}
\label{passsecprot}

In the previous section, we saw that Scheme A is prevented from fulfilling requirements for passive security by Charlie's knowledge of the parity of Alice and Bob's input bits. Here, we extend the protocol described in section ~\ref{secprot} and enhance it in order to make it passively secure, that is, secure in the case where party members create coalitions and are allowed to share their data, but still do not deviate from the protocol. 
%
It is clear that we must modify the protocol, such that Charlie never learns the value of the parity of $P_{i}$ and $Q_{i}$ bits. Initially this seems problematic, since Charlie needs this parity information to perform the needed measurement, as described in section~\ref{quantumnand} . We can avoid this, if we allow Alice and Bob to prepare the entangled states used in a special way, ``padding'' them with additional random Hadamard tranformations $H = \frac{\ket{0}+\ket{1}}{\sqrt{2}}\bra{0}+\frac{\ket{0}-\ket{1}}{\sqrt{2}}\bra{1}$ known only to themselves. If the parity bit received by Charlie is similarly padded, he can perform the required measurement on this state without ever learning the parity value.

The protocol thus proceeds as follows:

\begin{enumerate}
\item Repeat steps $2$-$11$ for all terms present in equation  \eqref{vandam}, starting with $i=1$.
\item Alice and Bob will calculate $P_i$ and $Q_i$ locally.
\item Alice and Bob each generate a local random ``preparation pad'' bit, $p_{a}$ and $p_{b}$.
\item Acting simultaneously, the three parties cyclically permute their qubits. Charlie gives his qubit to Alice, Alice hers to Bob and Bob gives his to Charlie. 
\item If $p_{a}=1$, Alice applies a Hadamard to Charlie's original qubit.
\item They cyclically permute the qubits again, Alice to Bob to Charlie to Alice. 
\item Now Bob possesses Charlie's original qubit. If $p_{b}=1$, Bob applies a Hadamard to this  qubit.
\item  They cyclically permute the qubits again, and each regains their initial qubit. (Note that the scheme can be amended such that no quantum communication is needed during the protocol -- see below.)
\item The state held by the parties is now $\openone\otimes\openone\otimes H^{p_{a}\oplus p_{b}}\ket{\psi}$. 
\item Alice and Bob privately give Charlie the bit values $P_{i}\oplus p_{a}$ and $Q_{i}\oplus p_{b}$ respectively.
\item Alice, Bob and Charlie measure their qubits in bases according to bit-values, $P_{i}$, $Q_{i}$ and $P_{i}\oplus Q_{i}\oplus p_{a}\oplus p_{b}$. 
\item When all $i$ terms are completed, Alice and Bob sum their local measurement outcomes. They send their summation bits to Charlie, who sums them with his own measured outcomes. 
\item Charlie reveals the value of $f(\vec{x},\vec{y})$.\end{enumerate}

This scheme can be shown to be passively secure with a threshold $t=2$. 

To see that the scheme produces the correct output note that Charlie's measurement is in the basis $H^{P_{i}\oplus Q_{i}\oplus p_{a}\oplus p_{b}}ZH^{P_{i}\oplus Q_{i}\oplus p_{a}\oplus p_{b}}$. This measurement is equivalent to first applying the operator $H^{p_{a}\oplus p_{b}}$, and thus effectively undoing the extra Hadamards applied by Alice and Bob,  and then performing a measurement determined via bit-value ${P_{i}\oplus Q_{i}}$. Thus the output of the protocol is equivalent to scheme A.

We shall show, in the next section, that the scheme is passively secure with threshold $t=2$. The scheme, as described above, has the undesirable feature that Charlie's qubit needs to be transmitted coherently between Charlie, Alice and Bob. However, this is not necessary. Instead, the parties could prepare an ensemble of  five-qubit states:

\begin{equation}
\rho=\sum_{{p_{a},p_{b}}} |p_{a}\rangle \langle p_{a}| \otimes |p_{b}\rangle\langle p_{b}| \otimes \left(\openone\otimes\openone\otimes H^{p_{a}\oplus p_{b}}\ket{\psi}\bra{\psi}\openone\otimes\openone\otimes H^{p_{a}\oplus p_{b}}\right)
\end{equation}

The first qubit is held by Alice, the second by Bob, the latter three by Alice, Bob and Charlie respectively. Instead of generating preparation pads, Alice and Bob simply measure their first qubits in the computational basis and use the outcomes as their pad bits. The state of the remaining qubits is then already $\openone\otimes\openone\otimes H^{p_{a}\oplus p_{b}}\ket{\psi}$. This replaces \textit{steps 3-8} of the protocol and Alice, Bob and Charlie can continue the protocol from \textit{step 9}. 

As will be proven in the next section, this scheme is passively secure with threshold $t=2$. The scheme is not actively secure, since in the final stage, Alice, Bob and Charlie can each change the final value reported by Charlie by lying in the final stage. We shall describe, in the discussion, some possible ways in which active security may be achieved.

\section{Passive Security Analysis of Scheme B} \label{actsecan}

In this section, we shall analyse scheme B step-by-step, considering for each, the information which each party learns about the others private data, and the information which they will gain if they form a coalition. We assume at each point that the parties follow the protocol precisely, thus any cheating is restricted to additional (classical) communication between corrupted parties. This is the standard setting for passive security in the SMPC literature.

In \textit{steps 1-3} no data is shared, thus no information can be learned by the parties in any case. In \textit{steps 3-9} the three parties cyclically permute their qubits, and Alice and Bob apply local transformations dependent on their private data. Since we are assuming that the protocol is followed by all parties, they may not measure their qubits. Even if they did make measurements, the local state of each qubit is maximally mixed and no information can be gained from the measurement.  It is important that at no time does any party possess two qubits, since, after Alice or Bob have applied their pad-Hadamard, a joint measurement of the padded qubit together with one other qubit can  reveal the pad bit. The reason for this is that the full stabiliser set of the GHZ state  $\ket{\psi}$ contains bi-partite operators such as $-\openone\otimes Y \otimes Y$, which transforms to $+\openone\otimes Y \otimes Y$ when a Hadamard is applied the second or third qubit. Any applied Hadamard would thus be  detectable via a measurement of  $\openone\otimes Y \otimes Y$. Furthermore, this attack would be hidden to the other party since it does not change the state. In the passive model, we assume that coalitions do not have the power to perform joint measurements, since that would require quantum communication between parties, which is considered an active deviation from the protocol. Thus, in this model, such an attack is disallowed. 

In \textit{step 10}, Charlie receives the bit-values $P_{i}\oplus p_{a}$  and $Q_{i}\oplus p_{b}$. From this information he obtains neither the values of $P_{i}$, $Q_{i}$ nor their parity. In order to obtain this, he would need to obtain the values of $p_{a}$ and/or $p_{b}$. He cannot obtain these values in the previous round, and his sole qubit at this stage is, from his perspective maximally mixed. Also, he cannot obtain these values by forming a coalition, which would, at best, provide him the private data of the coalition partner.
In \textit{step 11}, Alice, Bob and Charlie measure their qubits, and learn bit-values whose parity encodes the product $P_{i}Q_{i}$. This is an example of a ``shared secret``. All three parties must come together to learn the value of $P_{i}Q_{i}$ so this step is again secure, even under coalitions of 2 parties.
In \textit{steps  12 and 13} Alice and Bob sum their measured bits and send them to Charlie, who then announces the value of  $f(\vec{x},\vec{y})$. This is secure, even under passive coalitions, for the same reason as \textit{step 11}.

\textit{EPR-type attacks:} EPR-type attacks (named for the seminal Einstein-Podolsky-Rosen paper) \cite{Lo} have a special significance for passively secure quantum SMPC protocols as if a corrupted party can infer information about the honest parties data without being caught, using a quantum computer and delaying measurements, the role of passive security would be of reduced value in the context of quantum systems. This is because essentially the private data of honest parties would leak while the corrupted party is just performing local operations.

EPR attacks have been demonstrated in the case of $1$-sided protocols and while our protocol as presented in section ~\ref{passsecprot} is $n$-sided,  a $1$-sided variant can be easily created if \textit{steps 11,12} are modified as follows:

\begin{enumerate}
\setcounter{enumi}{10}
\item When all $i$ terms are completed, Alice and Bob sum their local measurement outcomes. Alice sends her summation bit to Charlie, who sums it with his own measured outcomes. 
\item Charlie sends the parity of his summed bits to Bob, who calculates the value of $f(\vec{x},\vec{y})$.\end{enumerate}

The question in such attacks is, can Bob learn the function outcomes for many values of his vector $\vec{y}$ without someone noticing? Bob is allowed to perform any quantum operation on his qubit(s) while he is attempting to infer the value of $f$ for Alice's given $\vec{x}$ and many possible $\vec{y}$'s. As proven by Lo in \cite{Lo}, if the entire protocol can be modelled as a unitary black box, then this attack successfully allows Bob to break the protocol,  
by applying unitary transformations to his part of the Hilbert space, which allow him to ``poll'' the black box for the output of the function  for many input vectors, and hence learn information about Alice's input. 

The reason why this attack fails in our protocol is that the repeated polling by any party is impossible, parties commit to an input value in two ways: first in the classical bit sent to Charlie in step 10, and secondly in the unbiased nature of the measurements corresponding to different input values. This means that consecutive polling by any corrupted party is impossible. 

If, on the other hand, corrupted parties were allowed to communicate quantumly, Bob could, for example, send his qubit to Charlie. Now Charlie could poll both possible input values. In possessing both qubits, he could make a joint measurement, and the relevant joint measurement pairs ($X\otimes X, Z\otimes Z$ or $X\otimes Z, Z\otimes X$) commute. Thus if quantum communication were allowed between corrupted parties, the EPR attack would succeed.

This feature is related to the property that GHZ type paradoxes occur in tri-partite but not in bi-partite systems, and illustrates that it is the inability to model the quantum passive secure model via a unitary black-box which is the key to avoiding the no-go theorem.

\section{No Quantum Cheating Channel (NQCC) Security Model}
\label{nqccmodel}

 We have now shown Scheme B to be passively secure. However, there is a problem with the notion of pure passive security in the quantum case. Assuming all parties are perfectly honest provides bit commitment for free and in combination with the results by Yao \cite{yaobitcom}, where it is proven that quantum bit commitment provides oblivious transfer, and Kilian \cite{kilianbitcom}, where is it proven that classical oblivious transfer provides classical SMPC, the definition of passive security itself would imply SMPC. Therefore, we expand our notion of security to include the case where corrupted parties are allowed to lie.

 In this section, introduce the notion of No Quantum Cheating Channel (NQCC) security model, where no restriction is imposed on the corrupted parties but the use of a quantum channel. We will consider a protocol to be NQCC compliant if, on top of safeguarding the honest party members data, it can detect attempts to corrupt the procedure, therefore allowing the execution of the protocol to be terminated.
 
 The characteristics of NQCC security are summarized in the following table,

\begin{center}
  \begin{tabular}{ | c || c |}
    \hline
    Property &  NQCC security \\ \hline
    Private data are not inferred by corrupted parties & \tick \\ \hline
    Corrupted parties are \textit{allowed to lie} & \tick \\ \hline
    Lying is detected & \tick \\ \hline
    Corrupted parties exchange classical information & \tick \\ \hline
    Corrupted parties exchange quantum information & \notick \\ \hline
 
  \end{tabular}
\end{center} 

The NQCC model is the most general security model, in a measurement-based scheme, which remains compliant with Lo's no-go theorem. Removing the only restriction imposed by this model, that is allowing the use of quantum channels, would make the system equivalent to a unitary Black box and then the attack invented by Lo compromises security. The value of NQCC security is that it sheds light on which part of Lo's theorem are the most crucial for measurement based SMPC. By highlighting these parts of the theorem, it could be possible that physical systems can be devised that are compliant with NQCC restrictions. These physical systems would then consist candidates for implementing secure multi-party computation at the quantum level.

\section{Scheme C: NQCC Secure Protocol}
\label{nqccprot}

In this Section, we extend Scheme - B, so that it becomes compliant with the NQCC security model. Since in the quantum case, the definition of passive security automatically implies SMPC, this extension is essential for the usefulness of our protocol. Furthermore, since NQCC is the most general form of security, which remains consistent with Lo's theorem, achieving this security level makes our protocol maximally secure under the restrictions imposed by quantum mechanics.

 Since Lo/EPR type attacks are not possible, due to restrictions imposed by the model, corrupted parties cannot learn the honest member's data but they can try to misinform him about the output by providing $1\oplus L$, where $L$ is the sum of their local measurements, during step $12$ of scheme B. If they can do so successfully, then the honest parties learn a false value of the outcome but the dishonest party member will learn the correct outcome.

In Scheme-C this will be detected as follows, if one party member, e.g. Alice artificially sets all her $P_i$'s equal to zero, $P_i = 0$, then the outcome of the function has to be zero. If Bob or Charlie are were bit flipping the sum of their local measurements, this would be revealed, as the function outcome would be non-zero.


 Therefore, parties repeat Scheme B many times, and in each execution of the protocol Alice and Bob would have a probability according to which they set all their $P_i$'s and $Q_i$'s respectively equal to zero. Instead of privately giving to Charlie the sum of their measurements, they announce it and along with that they announce if they were measuring with as security testers. Since many repetitions of a Scheme B are required, in order to detect cheating, we introduce one more index, $j$, which enumerates repetitions of Scheme B.
 
 Scheme-C can be summarized in the following steps:
 
 \begin{enumerate}
\item Agree on a number of repetitions, $N_{\text{rep}}$, Alice and Bob choose their probabilities to act as security testers, $t_a < 0.5$ and $t_b < 0.5$ respectively. Alice and Bob may choose the probabilities $0 < t_a <1 $, $0 < t_b < 1$ during the execution of protocol. 
\item For $j=1$, to $j=N_\text{rep}$ repeat the following steps:
\item Repeat steps $4$-$13$ for all terms present in equation  \eqref{vandam}, starting with $i=1$.
\item Alice and Bob will calculate $P_i$ and $Q_i$ locally. According to the values of $t_a$ and $t_b$ they may choose to set $P_i = 0$, $Q_i = 0$.
\item Alice and Bob each generate a local random ``preparation pad'' bit, $p_{a}$ and $p_{b}$.
\item Acting simultaneously, the three parties cyclically permute their qubits. Charlie gives his qubit to Alice, Alice hers to Bob and Bob gives his to Charlie. 
\item If $p_{a}=1$, Alice applies a Hadamard to Charlie's original qubit.
\item They cyclically permute the qubits again, Alice to Bob to Charlie to Alice. 
\item Now Bob possesses Charlie's original qubit. If $p_{b}=1$, Bob applies a Hadamard to this  qubit.
\item They cyclically permute the qubits again, and each regains their initial qubit. (Note that the scheme can be amended such that no quantum communication is needed during the protocol -- see below.)
\item The state held by the parties is now $\openone\otimes\openone\otimes H^{p_{a}\oplus p_{b}}\ket{\psi}$. 
\item Alice and Bob privately give Charlie the bit values $P_{i}\oplus p_{a}$ and $Q_{i}\oplus p_{b}$ respectively.
\item Alice, Bob and Charlie measure their qubits in bases according to bit-values, $P_{i}$, $Q_{i}$ and $P_{i}\oplus Q_{i}\oplus p_{a}\oplus p_{b}$. 
\item When all $i$ terms are completed, Alice, Bob and charlie sum their local measurement outcomes. They all concurrently announce their summation bits and also announce if they were acting as security testers for the current $j$.
\item Everyone calculates the value of $f(\vec{x},\vec{y})$.
\item If either Alice or Bob (or both) announced that they acted as security testers and $f(\vec{x},\vec{y}) \neq 0$, parties halt the protocol. An attempt to compromise it is detected.
\item If neither Alice nor Bob announced that they acted as security testers and the value of $f(\vec{x},\vec{y})$ is inconsistent with values for previous $j$'s when, again both were not security testers, the protocol is halted and an attempt to compromise it is detected.
\end{enumerate}
 
 Again, the qubit-swapping, can be avoided if instead of a GHZ state, the following five-qubt ensemble is shared between the three parties:
 
\begin{equation}
\rho=\sum_{{p_{a},p_{b}}} |p_{a}\rangle \langle p_{a}| \otimes  |p_{b}\rangle\langle p_{b}| \otimes \left(\openone\otimes\openone\otimes H^{p_{a}\oplus p_{b}}\ket{\psi}\bra{\psi}\openone\otimes\openone\otimes H^{p_{a}\oplus p_{b}}\right).
\end{equation}

\section{NQCC Security Analysis of Scheme C} \label{activesecuran}

Here, we examine Scheme-C and prove that it is compliant with NQCC security, that is the private data of honest parties remain uncompromised and cheating attempts are detected. The only assumption we will make is that no quantum channel is used.

Since the case of honest but curious corrupted parties was discussed in section ~\ref{actsecan}, here we will focus on attacks where corrupted parties deviate from the protocol. Except entering their private data feed to the protocol, parties dynamically interact (provide input) with the protocol, at \textit{steps} $7,9$, \textit{step} $12$ and \textit{step} $14$. The attacks which can be generated in these steps are the following:

\begin{enumerate}
\item{Alice or/and Bob provide to Charlie invalid bit values for $P_{i}\oplus p_{a}$ or $Q_{i}\oplus p_{b}$}
\item{Alice or Bob or Charlie Lie about the sum of their measurement bits.}
\item{Be dishonest on whether they acted as security testers}
\end{enumerate} 

\textit{Provide to Charlie invalid bit values for $P_{i}\oplus p_{a}$ or $Q_{i}\oplus p_{b}$}: In this attack, either Alice and/or Bob lie to Charlie and use fake preparation pads $p_a$ and/or $p_b$ respectively, which would lead Charlie to use an incorrect measurement axis. In this case, when Charlie is measuring along an incorrect axis, Charlie's measurement outcome would be $1$ with $50\%$ probability and $0$ with $50\%$ probability. This, is something detectable during \textit{step} $17$, as there will be inconsistency in the calculated function values between different protocol runs. 

\textit{Lie about the sum of their measurement bits}: Here, one party member is providing a bit flipped sum of his local measurements. This leads the other party members to learn a wrong (bit-flipped) function value, while the corrupted member would be able to recover the correct value. This compromising strategy has to be followed in every run of the protocol, else due to \textit{step} $17$, it will be detected. However, if, without loss of generality, Bob is performing this attack, he will be detected when Alice acts as a security tester at a run, during which he doesn't have the role of a security tester. This detection has a probability to happen on a repetition $j$ of the protocol equal to $t_a\left(1-t_b\right)$ and it happens on average after $\frac{1-t_b}{t_a}$ steps.

\textit{Be dishonest on whether they acted as security testers}: Here, a party member can either act as a security tester, enforcing a $f=0$ output, without announcing it or they could announce they acted as security testers without having set their input equal to zero. If the correct function outcome is $f=0$, this attack does not affect by any means the protocol and if the correct outcome is $f=1$ it will be detected during \textit{step} $16$.   
  
  Scheme C, is therefore NQCC secure and the security threshold, $t=2$, remains the same, since the detection methods work as long as all party members are not corrupted.
  
\section{Advantage over classical schemes}
\label{betterthresh}

So far, we have presented a scheme for two-party computation which is passively secure with threshold $t=2$. Compared to classical schemes, our scheme has the disadvantage that quantum resources and an extra player are needed. However,  by scaling up the scheme to multi-party computation over $n$ parties, a significant advantage of the quantum scheme is revealed.

Known passively secure classical schemes \cite{BGW88,CDDP99,BH06,DamsNiel} require, in the general case, an honest majority, in other words their threshold has an upper bound $t<n/2$. By modifying our scheme B, we can construct a scheme which allows secure multipartite computation over a restricted family of $n$-party functions with a threshold at its maximum value, $t=n-1$. The family of functions we consider are most easily described by considering $f(\vec{x},\vec{y})$ as a polynomial over $\mathbbm{Z}_{2}$ (i.e. where AND represents multiplication and XOR represents addition). They are polynomials of degree $2$ which have the following form.
 \begin{equation} 
f =  \bigoplus_{j_1 > j_2} \bigoplus_i \lambda_{j_{1},j_{2}}P_i^{j_1} P_i^{j_2}
\label{advantagedecomp}
\end{equation}
\noindent where the $j_1, j_2$ indices are used to distinguish the $n$ parties and $\lambda_{j_{1},j_{2}}$ is a bit number which indicates if a pair of parties $\left(j_{1},j_{2}\right)$ has a joint computation which is needed in order to evaluate $f$. As far as we are aware, there is no proof that a classical protocol for secure computation of degree 2 functions requires an honest majority, however our quantum protocol provides the highest possible corruption threshold. There exist recent examples (e.g. voting) of limited classical SMPC protocols which do not require an honest majority ~\cite{classvote1}. We hope that our result motivates more work in this area. 
 
Each term in the sum depends on input bits from 2 parties only, and hence scheme B can be adapted to provide a fully secure computation method. Notably, the threshold for this scheme will remain $n-1$.

The scheme progresses as follows

\begin{enumerate}
\item Repeat steps $2$-$3$ for each ($j_{1}$,$j_{2}$) term in equation  \eqref{advantagedecomp}. 
\item Parties $j_{1}$  and $j_{2}$ nominate a third party $j_{3}$. These three parties share a GHZ state.
\item The three parties follow Scheme B up to step 11. The parties retain their measured bits which are not yet shared.
\item After this has been completed for every participating pair, each party computes the parity of their measured bits, and announces the sum.
\item The players compute function $f$ by summing these public bits.
\end{enumerate}

In the $n$ party scheme, the security of the whole computation depends on the security of each of the three party computations performed. Since, at each stage prior to the final announcements, each term in the sum \eqref{advantagedecomp} is encoded in the parity of bits held by parties $j_{1}$, $j_{2}$ and $j_{3}$, each parties input data remains secure, even if all other $n-1$ other parties share data. For this reason, this scheme is passively secure with a threshold of $t=n-1$. Furthermore, if instead of Scheme B, the variant discussed in section~\ref{activesecuran} is used, then the scheme is NQCC secure, again with $t=n-1$.

\section{Discussion and conclusions}
\label{discuss}

In this paper we have introduced a scheme for secure $n$-party computation. The scheme exploits the intrinsically quantum correlations of the Greenberger-Horne-Zeilinger states to provide security and privacy. By considering a novel security model, inspired by the passive secure settings so important in classical secure computation, we show that unconditionally secure multi-party computation may be enhanced by access to  a quantum resource.
The scheme we present depends upon the natural secret-sharing characteristics of GHZ correlations.   Illustrating the potential of such correlations for private computation with a ``nearly private scheme'', we  the upgraded this nearly private scheme to a scheme (Scheme B) which is secure under the conditions of quantum passive security defined in section V. Afterwards, the protocol was further upgraded to NQCC security, where the parties are allowed to deviate from the protocol, as long as they do not use quantum communication.
This was then extended to a scheme for secure $n$-party computation with a threshold of $n-1$. This $n$-party scheme is restricted to quadratic functions, but achieves a security threshold higher than any known classical scheme.


The (non-physical) bi-partite object with analogous correlations to the GHZ state \cite{danclascom} is the Popescu-Rohrlich non-local box \cite{popescurohrlich}. Thus the non-local box (if it existed) would have a further application for secure computation. This observation was made independently very recently \cite{kaplanetalpaper} and used  to calculate better bounds on the number of Oblivious Transfer calls needed for secure computation of a function previous estimates \cite{BM04}.

Since the block on quantum communication between cheating players seems the key assumption which allows the security models we discussed to differentiate from the unitary black box model where the Lo-Colbeck no-go theorems apply, in order to physically realize quantum SMPC, models which fulfill this assumption need to be further researched. For example, noisy quantum storage models, which have recently beens shown to have some favourable cryptographic properties \cite{whehnerqmem}. 

One limitation of the scheme is its restriction to degree $2$ polynomials. Nevertheless, even for this restricted class of functions there is no known classical secure scheme which does not require an honest majority. Proving upper bounds on the security of classical schemes for restricted functions would be an interesting research direction, in which we are not aware that any work has been carried out. A recent generalisation \cite{ournewpaper} of \cite{danclascom} to higher degree functions may provide the means to extend our scheme to higher degree functions. It is possible that other families of functions with particular symmetries and structure are well suited to this kind of method. A further limitation, is the restriction to families of functions, which when written in the form of (\ref{advantagedecomp}), have a number of terms polynomial in the input size. This appears to be a fundamental limitation of employing the inner-product decomposition \cite{vandamssnloc}, since it can be shown that certain functions (e.g. the equality of two bit-strings) require exponentially many terms (see \cite{vandamssnloc} for a fuller discussion).


It is natural to ask whether the schemes we have presented can be developed into schemes for secure quantum computation, using cluster states \cite{onewayqc1} in place of the GHZ states. In fact, it has already been shown that cluster state-based quantum computation has promising security features, since a secure method of ``blind quantum computation'' \cite{blindcomputation} has been developed which utilises on cluster state measurement-based quantum computation. In light of this, the application of cluster states to secure quantum multi-party computation seems a promising direction.


The development of quantum key distribution has been one of the most successful aspects of quantum information science and is certainly the aspect closest to real-world application.  We hope that this work demonstrates that quantum methods can provide advantages in other cryptographic problems, and inspires further study in this area. 

\begin{acknowledgements}
We would like to thank Daniel Gottesman and Joe Fitzsimons for valuable suggestions. Also we are thankful to Brendon Lovett, Earl Campbell and Matthew Hoban for helpful discussions. This work was supported by QIPIRC and the National Research Foundation and Ministry of Education, Singapore. K. Loukopoulos acknowledges financial support from Materials Department, University of Oxford and St. Edmund Hall.
\end{acknowledgements}



\begin{thebibliography}{10}
\bibitem{footnote1}{One way to categorize SMPC protocols is through the number of parties that learn the computation outcome. If 1 party learns the outcome, then the protocol is called $1$-sided, if two parties learn the outcome, then the protocol is 2-sided and more generally, if $n$ parties learn the outcome, then the protocol is n-sided. Our protocol is presented as an $n$-sided protocol, however the other variations may be obtained by minor modifications, and we therefore do not emphasise this categorisation here.}
\end{thebibliography}
\end{document}